\documentclass[twocolumn,superscriptaddress,aps,showpacs,floatfix,prl,10pt]{revtex4-1}
\usepackage[T1]{fontenc}
\usepackage{amssymb,amsmath}
\usepackage{graphicx}
\usepackage{color}
\usepackage{bbold}
\usepackage{soul}
\usepackage[dvipsnames]{xcolor}
\usepackage{ulem}
\newcommand{\expect}[1]{\langle #1 \rangle}
\newcommand{\ket}[1]{| #1 \rangle}
\newcommand{\bra}[1]{ \langle #1 |}

\begin{document}

\title{Quantum reservoir processing}
\author{Sanjib Ghosh}
\email{sanjibghosh87@u.nus.edu}
\affiliation{School of Physical and Mathematical Sciences, Nanyang Technological University 637371, Singapore}
\author{Andrzej Opala}
\affiliation{Institute of Physics, Polish Academy of Sciences, Al. Lotnik{\'o}w 32/46, PL-02-668 Warsaw, Poland}
\author{Micha{\l} Matuszewski}
\affiliation{Institute of Physics, Polish Academy of Sciences, Al. Lotnik{\'o}w 32/46, PL-02-668 Warsaw, Poland}
\author{Tomasz Paterek}
\affiliation{School of Physical and Mathematical Sciences, Nanyang Technological University 637371, Singapore}
\affiliation{MajuLab, CNRS-UCA-SU-NUS-NTU International Joint Research Unit, UMI 3654, 117543 Singapore}
\author{Timothy C. H. Liew}
\email{timothyliew@ntu.edu.sg}
\affiliation{School of Physical and Mathematical Sciences, Nanyang Technological University 637371, Singapore}
\begin{abstract}
\textbf{The concurrent rise of artificial intelligence and quantum information poses opportunity for creating interdisciplinary technologies like quantum neural networks. 
Quantum reservoir processing, introduced here, is a platform for quantum information processing developed on the principle of reservoir computing that is a form of artificial neural network. 
A quantum reservoir processor can perform qualitative tasks like recognizing quantum states that are entangled as well as quantitative tasks like estimating a non-linear function of an input quantum state (e.g. entropy, purity or logarithmic negativity). In this way experimental schemes that require measurements of multiple observables can be simplified to measurement of one observable on a trained quantum reservoir processor.} 
\end{abstract}
\maketitle

\section{Introduction}

Quantum neural networks are emerging technologies that combine the features of artificial neural networks and quantum information technologies \cite{Biamonte17,Dunjko18,Altaisky16,Adcock15}. While neural networks are biologically inspired computing systems that learn from example to perform complex tasks in the area of ``big data'' and machine learning \cite{Stajic15,Chouard15,LeCun15,Butler18}, quantum information technologies exploit quantum effects for practical applications like quantum computation, quantum cryptography and long distance quantum communications. The interaction between these two promising fields led to many advances. For instance, quantum effects in neural networks \cite{Lewenstein94,Kak95} enhance learning efficiency \cite{Dunjko16,Wiebe16} and speed-up solving many classical tasks \cite{Neigovzen09, Benedetti16, Alvarez-Rodriguez17}. Conversely, neural networks are used for solving complex quantum problems \cite{Carleo17,Behler07} and the control and design of quantum experiments \cite{Krenn16,Melnikov18,Bukov18}.

Among the forms of neural networks, recurrent neural networks emerged as particularly suited for solving complex temporal machine learning tasks. They achieve this by using feedback connections not present in more traditional feedforward neural networks to generate an internal temporal dynamic behaviour. However, the training of recurrent neural networks is typically inefficient and computationally expensive.

In reservoir computing, a randomly connected network, called the reservoir, is used as a dynamical computing unit into which an input signal is fed. 
The training in reservoir computing takes place only at the readout weights that linearly maps the readout of the reservoir state to the desired output. The training is conceptually simple and computationally inexpensive \cite{Lukosevicius12}. Apart from these advantages, they are very suitable for hardware implementation in a wide variety of systems \cite{Paquot12,Larger12,Duport12,Brunner13,Vandoorne14,Du17,Kudithipudi16,Larger17,Tanaka18}. Despite these advantages, reservoir computing is mostly used for tasks in the classical domain, like time series prediction and speech recognition \cite{Maass02,Brunner13,Larger17,Vandoorne14}, predicting the evolution of nonlinear dynamics \cite{Jaeger04} and features of chaotic systems \cite{Pathak18}.

\begin{figure}[h]
\includegraphics[width=1\columnwidth]{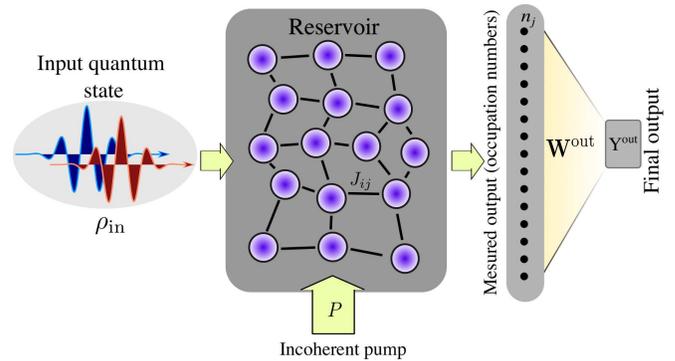}
\caption{\textbf{Schematic representation of a quantum reservoir processor.} A quantum state in the form of an optical field excites a fermionic lattice with random coupling $J_{ij}$ in an effective Fermi-Hubbard model. The occupation numbers of the fermionic sites are extracted and combined to give a final output. 
This generic architecture can perform various tasks, such as identifying a quantum state and simultaneously estimating its various properties. }
\label{QuantumReservoirComputer}
\end{figure}

\begin{figure}[hht]
\includegraphics[width=1\columnwidth]{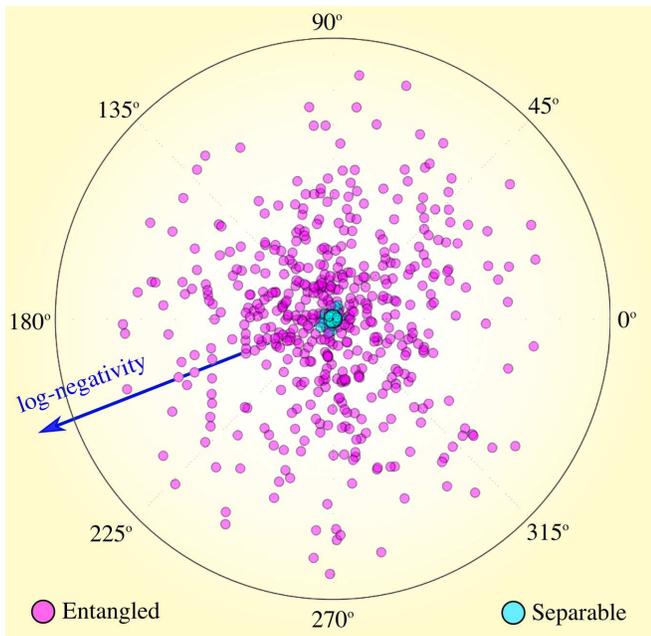}
\caption{\textbf{Quantum reservoir processor trained to recognize entangled squeezed-thermal states.} 
A reservoir processor of $4$ fermions was trained with $200$ squeezed-thermal states and tested with another set of squeezed-thermal states with squeezing parameter $|\alpha| e^{i\theta}$. 
Each point in the radial plot shows an input state $\rho_\text{in}$ with radius being the logarithmic negativity $\mathcal{N}(\rho_\text{in})$ and angle being the squeezing phase $\theta$. 
Clearly, the predicted separable states are largely concentrated at $\mathcal{N}(\rho_\text{in})=0$ and the entangled states are largely at $\mathcal{N}(\rho_\text{in}) > 0$. 
The overall prediction error is $(3.7 \pm 0.7)\%$. }
\label{Test_Squeezed_Thermal}
\end{figure}

Here we present a quantum reservoir processing platform using a quantum reservoir to perform quantum tasks on a quantum input. 
Specifically, we consider a 2D fermionic lattice with random intersite coupling excited by an incident quantum state in the form of an optical field, as illustrated in Fig.~\ref{QuantumReservoirComputer}. 
We find that this architecture is versatile and can perform both qualitative and quantitative tasks.
Recognition of quantum entanglement of the input state is an example of a qualitative task.
We find that the quantum reservoir processor (QRP) not only recognizes the entanglement of the same class of states as the training set but is also able to make predictions on states beyond the training class, including bipartite bound entangled states.
Our examples of quantitative tasks include estimation of logarithmic negativity, von Neumann entropy, purity and the trace of any power of an input quantum state.
We discuss consequences of these findings to simplification of generic quantum experiments.
In particular, we argue that measurements of multiple quantum observables can be replaced with a single measurement using QRP that has been suitably trained.

\section{Results}

\subsection{The model}

Our considered quantum reservoir is a set of fermions arranged in a 2D lattice with random nearest-neighbour hopping. The reservoir is defined by the Fermi-Hubbard Hamiltonian:
\begin{eqnarray}
\hat{H}_R =  \sum_{ij} J_{ij} \left(  \hat{b}_i^\dagger \hat{b}_j + \hat{b}_j^\dagger \hat{b}_i  \right)
\end{eqnarray}
where $\hat{b}_i$ is the fermionic field operator (spin less) of the site $i$ and $J_{ij}$ are the random hopping amplitudes uniformly distributed in the interval $[- \gamma,+\gamma]$, where $\gamma$ is the decay rate. 
Each site is driven by an incoherent excitation (e.g., non-resonant optical field) with the strength $P = 0.1\gamma$ \cite{Laussy08,Valle10}. 
In our scheme, an input bipartite state, in bosonic (e.g., optical) modes $\hat a_1$ and $\hat a_2$, is represented by the density matrix $\rho_\text{in}$. 
It is incident on the reservoir, interacting for a short time with all fermions.
We consider that the two modes of the input state are coupled to the reservoir one at a time. This is to model the physical process where wave packets are sequentially incident on the reservoir one after the other. The couplings of the input modes to the reservoir are realized via the ``cascaded formalism'' \cite{Carreno16} which eliminates any feedback from the reservoir to the input modes. Due to this coupling, the input state merges to the reservoir. The incident state thus influences the evolution of the reservoir. 
As readout, we measure the occupation number of each fermionic site of the reservoir.
This model can be practically realised in a variety of platforms, 
including arrays of semiconducting quantum dots or superconducting qubits~\cite{Georgescu14}. 
We note that precise and deterministic quantum dots, which are typically a key challenge~\cite{Schmidt07}, are unnecessary for our scheme where random positioning and coupling is actually useful.

The whole phenomenon can be described by the combined density matrix $\rho$ which includes the quantum reservoir and the incident modes. It follows the quantum master equation:
\begin{eqnarray}
i\hbar \dot{\rho} &=& [\hat{H}_R,\rho] + \frac{i\gamma}{2} \sum_j \mathcal{L}(\hat{b}_j)  + \frac{iP}{2} \sum_j \mathcal{L}(\hat{b}_j^\dagger) \nonumber \\
&+& i \, \sum_{k,j} f_k(t) W^\text{in}_j \left( [\hat{a}_k\rho, \hat{b}_j^\dagger ] +  [ \hat{b}_j, \rho \hat{a}_k^\dagger] \right) \nonumber \\
&+& \frac{i\eta} {2\gamma} \sum_k f_k(t) \mathcal{L}(\hat{a}_k)
\label{MasterEq}
\end{eqnarray}
where the last two lines realise the cascaded formalism~\cite{Carreno16}, with $\eta =\sum_j (W^\text{in}_j )^2$ 
and the functions $f_k(t)$ (for $k=1,2$) indicate that the input modes $\hat{a}_k$ are coupled to the reservoir for brief periods of time at different instances. 
Specifically, we consider that $f_1(t)=1$ for $t_1<t<t_1+\tau$ when the first mode $\hat{a}_1$ is connected to the reservoir, whereas $f_2(t)=1$ for $t_1+\tau<t<t_1+2\tau$, when the second mode is connected to the reservoir; both $f_{1,2}(t)=0$ at any other time. Time $t_1$ describes duration for the reservoir to reach the steady state. 
The Lindblad operator reads $\mathcal{L}(\hat{x}) = 2 \hat x \rho \hat x^\dagger - \hat x^\dagger \hat x \rho - \rho \hat x^\dagger \hat x$.
For our numerical simulations, we consider $\tau=\hbar/\gamma$ and the input weight matrix ${\bf W}^\text{in}$ with random components uniformly distributed in $[0,\gamma]$. 

The occupation numbers $n_j = \expect{\hat{b}_j^\dagger \hat{b}_j }$ of the reservoir fermionic sites provide a readout measured at $t=t_1+2\tau$. Our desired output can then be defined as $Y_i^\text{out} = \sum_j W^\text{out}_{ij} n_j $, that is, a linear combination of the readout occupation numbers. The output weight matrix $\bf{W}^\text{out}$ is optimized using a training data set such that the $Y^\text{out}$ is best fitted with known training data, corresponding to a particular task.
We will now describe exemplary tasks.

\subsection{Recognition of quantum entanglement} 
We first train the QRP with a set of bipartite squeezed-thermal states that are randomly distributed between separable and entangled states. 
The two-mode squeezing operator $\hat{\mathcal{S}}(\alpha) = \exp{ (\alpha \hat{a}_1^\dagger \hat{a}_2^\dagger - \alpha^* \hat{a}_1 \hat{a}_2)} $ is applied on bipartite thermal states $\rho_\text{th}$, with average occupation number per mode $\overline{n}$, to obtain the squeezed-thermal states:
\begin{eqnarray}
\rho_\text{in}  =  \hat{\mathcal{S}}(\alpha)  \,  \rho_\text{th}  \, \hat{\mathcal{S}}^\dagger (\alpha)
\end{eqnarray}
where the squeezing parameter $\alpha = |\alpha | e^{i\theta}$, and $|\alpha |$ and $\theta$ are chosen random such that on average $50\%$ of states are entangled while others are separable (see Supplementary Material). 
The task is to find the states that are entangled.

For the considered supervised training, the input states must be unambiguously classified into entangled and separable. The squeezed-thermal states are bipartite Gaussian states, thus can be unambiguously characterized by the logarithmic negativity \cite{Werner01}. We train the processor using a set of these states by assigning $Y^\text{out}=(1,0)$ if a state is entangled and $Y^\text{out}=(0,1)$ otherwise. The training determines the optimum output weights $\textbf{W}^\text{out}$ by minimizing the prediction error using ridge regression.

For a performance test, we again prepare a set of random input states $\rho_\text{in}$ which are then fed to the quantum reservoir processor.  For each input $\rho_\text{in}$, the processor then provides an output $\bf Y^\text{out}$, which is a 2D vector. If the first element of the vector is larger than the other element, then we assign the input state as entangled and otherwise as separable. 
In order to test the prediction efficiency, we calculate the logarithmic negativity $\mathcal{N}(\rho_\text{in})$ to independently verify whether $\rho_\text{in}$ is entangled. 
In an ideal situation, the processor predicts $\rho_\text{in}$ as entangled whenever $\mathcal{N}(\rho_\text{in}) >0$. In Fig.~\ref{Test_Squeezed_Thermal}, we represent the input states in a polar plot, where the radius is representing the log-negativity of the state and the angle representing the squeezing angle $\theta$. The prediction of the reservoir processor is presented by the color of each point. The predicted entangled states are the magenta points and the predicted separable states are the blue points. We can see that the separable states are clustered at the center of the polar plot indicating that the predicted separable states are of zero or extremely low log-negativity.

\begin{figure}[h]
\includegraphics[width=1\columnwidth]{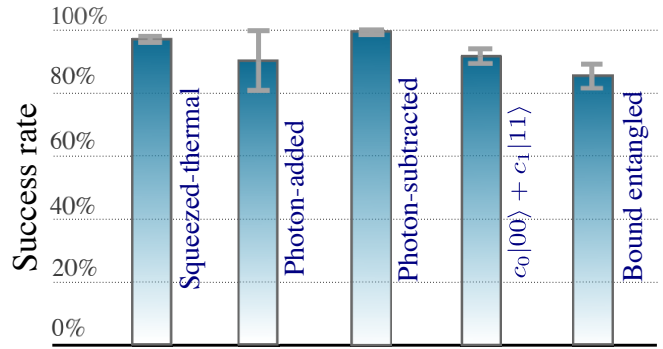}
\caption{\textbf{quantum reservoir processor recognizes other classes of entangled states.} 
In this simulation the QRP consists of $4$ fermions.
The heights of the bars give the percentage of sampled states with correctly identified separability properties.
The processor is trained with $200$ examples of only squeezed-thermal states. The data are averaged over $10$ different configurations of the random couplings $J$ between the fermions and input weights $\textbf{W}^\text{in}$, and the error bars are indicating the corresponding standard deviations.
}
\label{SuccessRateNonGaussian}
\end{figure}

It turns out that the separability criterion recognized by the QRP is applicable to a wider class of input states beyond the training states. 
We consider frequently used non-Gaussian states (see Methods for detailed expressions): two-mode squeezed states with a photon added or subtracted (mean photon number comparable to that in the training set), 
state $c_0 \ket{00} + c_1 \ket{11}$, and bound entangled states introduced in Ref.~\cite{Horodecki00}.
We emphasise that QRP is trained only with the squeezed-thermal states. 
Surprisingly, it recognizes the non-Gaussian entangled states very efficiently. 
This suggests that the processor has truly identified the entanglement pattern from the considered Gaussian input states and has used that pattern to recognize the non-Gaussian entangled states, see Fig.~\ref{SuccessRateNonGaussian}.

\begin{figure*}[]
\includegraphics[width=2\columnwidth]{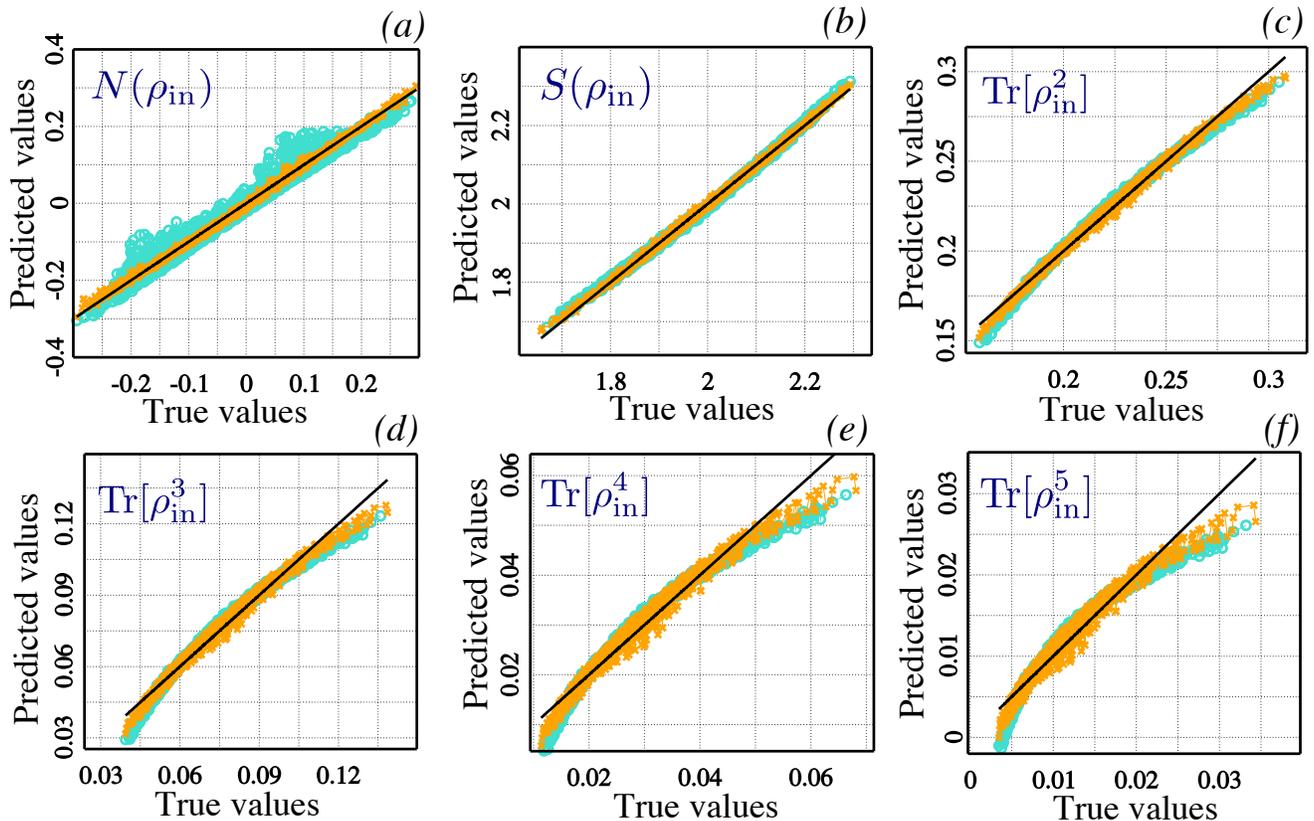}
\caption{\textbf{Quantitative predictions for non-linear functions of an input quantum state.} 
Here we demonstrate simultaneous estimation of six parameters.
In each panel, we plot the true values versus the predicted values with reservoirs of $2$ fermions (blue) and $4$ fermions (orange). 
The solid black line corresponds to the ideal predictions. 
The predicted values become more precise and more accurate with the increasing number of fermionic sites in the reservoir. 
Panel (a) is for logarithmic negativity (where we retain negative values), panel (b) for von Neumann entropy, and the remaining panels for trace of higher powers of the input state.}
\label{MultiObservables}
\end{figure*}

\subsection{Quantitative estimations and multiprocessing} 

QRP can also perform accurate quantitative estimations of non-trivial physical quantities.
Furthermore, the method allows simultaneous estimation of many parameters and observables.
Suppose we want to estimate $M$ quantities of interest given an input state $\rho_\text{in}$.
For this we take $ \textbf{Y}^\text{out}$ as an $M$ dimensional vector. 
In the training phase, each element of the output vector, $Y_i^\text{out}$, is taken as an estimate of the $i$th parameter.
Once the optimum output weight matrix $\textbf{W}^\text{out}$ is obtained from the training states, the QRP can predict the values of all $M$ parameters at once.

As an example consider the following set of six parameters: log-negativity (retaining the negative values, see Supplementary Material) $M_0 = N(\rho_\text{in})$,
von Neumann entropy $M_1 = S(\rho_\text{in})$, and $M_n = \text{Tr}(\rho_\text{in}^n)$ for $n = 2 \dots 5$.
Clearly any arbitrary parameter with series expansion $\langle \sum_n c_n \rho^n \rangle$ can be estimated similarly.
We again used the squeezed-thermal states as a training set in order to obtain the weight matrix $\textbf{W}^\text{out}$. 
Fig.~\ref{MultiObservables} shows the excellent capability of the QRP for predicting accurate and precise values of all parameters in one go. 
We see that estimation is better for bigger size of the quantum reservoir. 
Thus it is expected to have an (almost) perfect prediction capability when the reservoir size is large.

In this way QRP provides a universal platform for simplification of quantum experiments.
In a typical experiment a good estimation of a non-linear function of $\rho_\text{in}$ requires measurements of multiple quantum observables.
In the worst case one has to perform full quantum state tomography.
This is a consequence of the fact that the probability of measurement result $r$ is a linear function of $\rho_\text{in}$, i.e. $p_r = \mathrm{Tr}(\rho_\text{in} \Pi_r)$, where $\Pi_r$ is the corresponding POVM element.
The advantage of QRP is that only \emph{one} measurement is conducted (on the reservoir) and then different parameters are obtained by post-processing of the results.
This comes at the expense of additional resources needed to train the processor.
As seen in our example, the quality of prediction depends on the number of fermionic sites in the reservoir.
If the required precision is obtained with a QRP small enough to be simulated on a classical processor, 
the training can be done by supplying density matrices likely to be produced in experiment 
(or random mixed states in the case of no prior knowledge of the experiment).
For a large QRP, the training requires supplying well-characterised physical input states, for which the parameters of interest can be calculated independently and efficiently.
Note that this needs to be done only once.

\section{Discussion}
We have presented a quantum reservoir processing platform for recognition of quantum entanglement and estimation of non-linear functions of the input state.
This architecture can be used both as a programmable quantum hardware device that can be programmed by training according to the need, or as a software architecture for quantum machine learning that can work for quantum tasks which are otherwise hard. 

For instance, a software implementation could be to use a quantum reservoir processing platform for identifying bound entangled states. 
For hardware implementation, our considered reservoir, that is a 2D fermionic lattice, can be realized in a variety of systems, such as semiconductor quantum dots, NV centres in diamond and trapped atoms. The model of our reservoir could be equivalently realized with a driven-dissipative array of fermionized photons \cite{Carusotto09}, possibly using photonic crystal cavities \cite{Gerace09}. Exciton-polaritons in semiconductor microcavities offer yet another alternative, which are now approaching the polariton blockade regime \cite{Verger06, Munoz-Matutano17, Ningyuan18, Delteil18} and were shown to receive the entanglement of external optical fields \cite{Cuevas18}.

\section{Methods}
\textit{Squeezed-thermal states:-}
A bipartite thermal state can be represented by the density matrix: $ \rho_\text{th} =\sum_{n_1,n_2} \rho_{n_1n_2} \ket{n_1,n_2} \bra{n_1,n_2}$ with the Fock space elements:
\begin{eqnarray}
\rho_{n_1n_2} = \left( \frac{1}{ 1+ \overline{n} } \right)^2 \left( \frac{\overline{n} }{ 1+\overline{n} }  \right)^{n_1+n_2}
\end{eqnarray}
and $\overline{n}$ is the average occupation number per mode. The squeezed-thermal states are then obtained as $\rho_\text{sq-th}  =  \hat{\mathcal{S}}(\alpha)  \,  \rho_\text{th}  \, \hat{\mathcal{S}}^\dagger (\alpha)$ where the squeezing operator $\hat{\mathcal{S}}(\alpha) = \exp{ (\alpha \hat{a}_1^\dagger \hat{a}_2^\dagger - \alpha^* \hat{a}_1 \hat{a}_2)} $ and the thermal state $ \rho_\text{th}$ is charaterized by the average thermal occupation number $\overline{n}$. 
We write the squeezing parameter as $\alpha =|\alpha| e^{i\theta}$ and further $|\alpha | = s \sin\phi$ and the average thermal occupation number $\overline{n} = s^2 \cos^2\phi$. 
Thus, the parameters $\theta$, $s$ and $\phi$ are the parameters charaterizing the states $\rho_\text{sq-th}$. 
We take $\theta$, $s$ and $\phi$, as random numbers uniformly distributed in the intervals  $[0,2\pi]$, $[0.8,0.95]$ and $0.5 \pm\pi/10$, respectively. 
We have chosen the intervals for all the parameters such that $50\%$ of the states are Gaussian entangled.

\textit{Photon added squeezed states:-}
The photon added squeezed states are written as 
$\rho_\text{sq-add}  =  \hat{a}_1^\dagger  \hat{a}_2^\dagger  \hat{\mathcal{S}}(\alpha)  \, \ket{00}\bra{00}  \, \hat{\mathcal{S}}^\dagger (\alpha) \hat{a}_2 \hat{a}_1 $
where we have considered $\alpha =|\alpha| e^{i\theta}$ with $|\alpha|$ and $\theta$ uniformly distributed in $[0.1,0.25]$ and $[0,2\pi]$, respectively. 
The separability of states is not always easy to recognize. For example, the Simon criterion does not detect the entanglement of these states for $|\alpha| < 0.378$ \cite{Simon00,Nha12}.
We have chosen the parameter $\alpha$ in such a way that the prepared states have an average occupation number close to that of the training squeezed-thermal states. 

\textit{Photon subtracted squeezed states:-}
The photon subtracted squeezed states are experimentally relevant \cite{Ourjoumtsev09}. These states are expressed as
$\rho_\text{sq-sub}  =  \hat{a}_1  \hat{a}_2  \hat{\mathcal{S}}(\alpha)  \, \ket{00}\bra{00}  \, \hat{\mathcal{S}}^\dagger (\alpha) \hat{a}_2^\dagger \hat{a}_1^\dagger$ where we have considered $\alpha =|\alpha| e^{i\theta}$ with $|\alpha|$ and $\theta$ uniformly distributed in $[0.8,0.95]$ and $[0,2\pi]$, respectively. We have chosen the parameter $\alpha$ in such a way that the prepared states have an average occupation number close to that of the training states (squeezed-thermal). 

\textit{The states $c_0 \ket{00} + c_1\ket{11}$:-}
For these states, we have considered the parameterization $c_0 = \sin\theta$ and  $c_1 = \cos\theta \, e^{i\phi}$, and we have sampled these states uniformly on a Bloch sphere.

\textit{Bound entangled states:-}
We considered a family of  bound entangled states defined by~\cite{Horodecki00},
\begin{eqnarray}
\rho_\text{bn} = \frac{1}{A} \left(  \ket{\Psi} \bra{\Psi}   + \sum_{n=1}^\infty \sum_{m>n}^\infty \ket{\Psi_{mn}} \bra{\Psi_{mn}}   \right)
\end{eqnarray}
where $A$ is the normalization constant, $ \ket{\Psi} = \sum_{n=1}^\infty a^n \ket{n,n}$ and $\ket{\Psi_{mn}} = c^m a^n \ket{n,m} + a^m c^{-m}  \ket{m,n}$. 
The two parameters $a$ and $c$ satisfy the condition $ 0<a<c<1$ to impose finite $A$. 
$c$ and $a/c$ are chosen randomly in ranges $[0.3,0.6]$ and $[0+,0.1]$, respectively. These states are bound entangled in the infinite dimensional continuous variable limit as well as in finite dimensions~\cite{Horodecki00}. We achieve the continuous variable limit with a small $a/c <0.1$ in a truncated Fock space.

\section{Acknowledgments}
This work is supported by the Singapore Ministry of Education Academic Research Fund Tier 2, Project No. MOE2015-T2-2-034 and MOE2017-T2-1-001. MM and AO acknowledge support from the National Science Center, Poland grant No. 2016/22/E/ST3/00045

\newpage
\section{Supplementary material for ``Quantum reservoir processing''}

\begin{figure}[h]
\includegraphics[width=0.85\columnwidth]{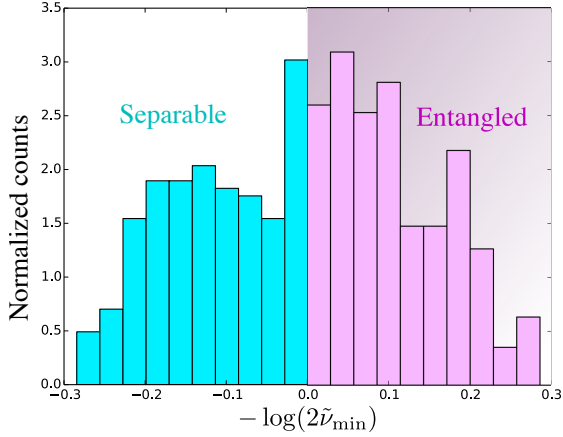}
\caption{Here we show the normalized distribution of the considered squeezed-thermal states $\rho_\text{sq-th}$ for different values of $N(\rho_\text{sq-th})=-\log(2\tilde{\nu}_\text{min})$. As indicated by the color of the bars, about $50\%$ of states are entangled (magenta) and others are separable (blue). The method for generating the squeezed-thermal states is described in the main article.}
\label{SqThDistribution}
\end{figure}

\textbf{Distribution of the squeezed-thermal states:-}
For training, we have used squeezed-thermal states. 
These states are generated using the parameters $s$, $\phi$ and $\theta$ that are introduced in the Methods section of the main article. 
We chose the specific ranges of variation for these parameters to ensure that $\sim 50\%$ states are entangled. 
This is shown in Fig.~\ref{SqThDistribution}. 
The equal division of the input states between entangled and separable sates avoids any bias during the training of the reservoir processor.

\begin{figure}[]
\includegraphics[width=0.85\columnwidth]{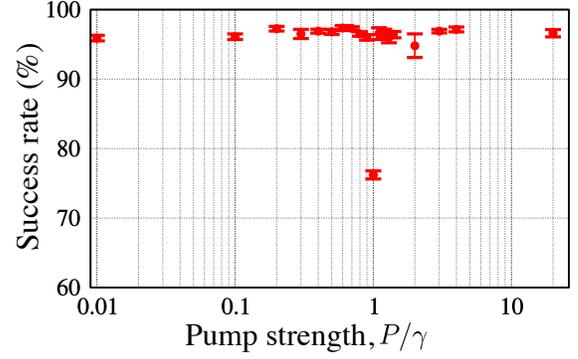}
\caption{Success rates for recognizing the separability of squeezed-thermal states as a function of the incoherent pump strength $P$. Here the readout is trained with $200$ random squeezed-thermal states. The data is averaged over $10$ random realizations of the reservoir consisting of $3$ fermions.}
\label{PumpFunction}
\end{figure}

\textbf{Function of the incoherent pump:-}
Here we show the effect of the incoherent pump (gain) applied to the reservoir. In Fig.~\ref{PumpFunction}, we show the performance of a quantum reservoir processor in recognizing separability of squeezed-thermal states for different pump strengths, $P$. Except for $P\approx \gamma$, the performance of the reservoir processor is very high for any pump strength. This means, no fine-tuning in $P$ is required for realizing the reservoir.  Moreover, the reservoir processor functions well even for $P=0$ with only $1\%$ reduction in the success rate. Thus, the incoherent pump could be totally removed from our scheme (experimentally significant) without any major change in the performance. 

\begin{figure}[]
\includegraphics[width=1\columnwidth]{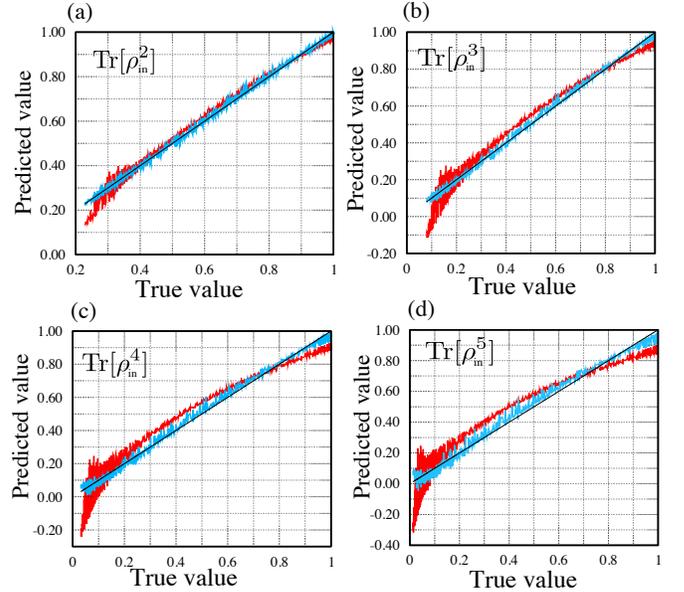}
\caption{Here we show the prediction of the quantum reservoir processor for a different set of parameters than what is considered in the main article such that the quantities $\text{Tr}[\rho_\text{in}^n]$ are distributed in the full range $[0,1]$. The panels (a)-(d) are showing the plots for true values versus the predicted values of $\text{Tr}[\rho_\text{in}^n]$ for $n=2 \dots 5$ respectively. In each panel, we plot the data for reservoirs with $2$ fermions (red) and $4$ fermions (blue) and the solid black lines corresponds to the ideal prediction. 
As seen the prediction becomes better with the increasing number of fermions in the reservoir. 
Here the parameter $s$, for generating the input squeezed-thermal states, is taken random in the range $[0,0.8]$ (see the main article for details). }
\label{Purity0to1}
\end{figure}

\textbf{Wider sampling of the training set:-}
Here we present the prediction of the quantum reservoir processor for a different set of parameters than those considered in the main article. Here the purity of the states varies in the full range $0$ to $1$. Thus sampling of the training input states is much wider. However, these input states are not equally distributed between entangled and separable states. In Fig.~\ref{Purity0to1}, we show the predictions of the quantum reservoir processor for $\text{Tr}[\rho^n_\text{in}]$ with $n=2,3,4,5$. With this wider sapling of the input states, we find that the prediction accuracy increases for $\text{Tr}[\rho^n_\text{in}]$ compared to the one presented in the main article. However, this training set is inefficient for entanglement recognition task.

\textbf{Entangled cat states:-}
We have seen some examples of entangled states that are recognized by the reservoir processor trained with only the squeezed-thermal states. Here is an example, entangled cat states, that are not recognized with the same procedure:
\begin{eqnarray}
\rho_\text{in} \propto &&\ket{\small{+\beta,+\beta}} \bra{\small{+\beta,+\beta}}  + \ket{\small{-\beta, -\beta}} \bra{\small{-\beta,-\beta}} \nonumber  \\
 &-& p \left(  \ket{\small{-\beta,-\beta}} \bra{\small{+\beta,+\beta}}  + \ket{\small{+\beta,+\beta}} \bra{\small{-\beta,-\beta}} \right)
\end{eqnarray}
where $\ket{\small{\pm\beta}}$ are coherent states and $p$ is the degree of coherence. 
However, when the processor is trained with the same class of entangled cat states with randomly chosen $p$, the processor shows an extremely high success rate (more than $90\%$).

\textbf{Truncated Fock space for the input continuous field:-}
For the input continuous field, the single particle creation operator can be written in matrix representation in the Fock space:
\begin{eqnarray}
a^\dagger = { \footnotesize \begin{pmatrix}           
0 & 0 & 0 & \dots & 0 &\dots \\
\sqrt{1} & 0 & 0 & \dots & 0 & \dots\\
0 & \sqrt{2} & 0 & \dots & 0 & \dots\\
0 & 0 & \sqrt{3} & \dots & 0 & \dots\\
\vdots & \vdots & \vdots & \ddots  & \vdots  & \dots\\
0 & 0 & 0 & \dots & \sqrt{n} &\dots &  \\
\vdots & \vdots & \vdots & \vdots & \vdots  &\ddots \end{pmatrix} }
\end{eqnarray}
We truncate the matrix at the occupation number $4$, such that the truncated creation operator is now represented as,
 \begin{eqnarray}
a^\dagger = {\footnotesize \begin{pmatrix}           
0 & 0 & 0 & 0 \\
\sqrt{1} & 0 & 0 & 0 \\
0 & \sqrt{2} & 0 & 0 \\
0 & 0 & \sqrt{3} & 0 \\
0 & 0 & 0 & \sqrt{4}  \end{pmatrix} }
\end{eqnarray} 
Accordingly, we can span the full Hilbert space in the truncated Fock space by stopping at the occupation number $4$ for all other operators. Note that the single particle Fermionic creation operator of a reservoir fermion can be written as 
\begin{eqnarray}
b^\dagger =\begin{pmatrix}           
0 & 0  \\
\sqrt{1} & 0  \end{pmatrix}
\end{eqnarray}
without any approximation.

\textbf{Logarithmic negativity:-}
A bipartite Gaussian state is described either by its density matrix $\rho_\text{in}$ or by the covariance matrix $V$ with elements
\begin{eqnarray}
V_{ij} = \frac{1}{2} \expect{ X_i X_j + X_j X_i }  -  \expect{X_j} \expect{X_i }
\end{eqnarray}
where $X = (q_1,p_1,q_2,p_2)$, $q_i = (\hat{a}_i+\hat{a}_i^\dagger) / \sqrt{2}$,  $p_i = (\hat{a}_i- \hat{a}_i^\dagger) / (i\sqrt{2})$ and $\hat{a}_i$ are the annihilation operators. The covariance matrix can be written as,
\begin{eqnarray}
V =
  \left( {\begin{array}{cc}
   A &{~~} C \\
   C^T & {~~}  B \\
  \end{array} } \right)
\end{eqnarray}
where each of $A, B$ and $C$ are $2\times 2$ matrices. The logarithmic-negativity of a Gaussian state is defined by,
\begin{eqnarray}
\mathcal{N}(\rho_\text{in}) = \text{max} \left[ 0, -\log ( 2\,\tilde{\nu}_\text{min} ) \right]. 
\end{eqnarray}
The logarithmic-negativity that retains the negative values reads,
\begin{eqnarray}
N(\rho_\text{in}) = -\log ( 2\,\tilde{\nu}_\text{min}) \\ \nonumber
\end{eqnarray}
where $\tilde{\nu}_\text{min} = \sqrt{ \Sigma-\sqrt{\Sigma^2-4\,\text{Det}V } } /\sqrt{2}$ is the smallest symplectic eigenvalue of the partially transposed covariance matrix and $\Sigma = \text{Det} A + \text{Det} B - 2\, \text{Det} C $. 
The state $\rho_\text{in}$ is entangled if $\mathcal{N}(\rho_\text{in}) >0$. One may work with the definition $q_i=(\hat{a}_i+\hat{a}_i^\dagger) $ and $p_i = (\hat{a}_i- \hat{a}_i^\dagger) / i$. Then the corresponding symplectic eigenvalue $\tilde{\nu}_\text{min}' = 2 \tilde{\nu}_\text{min} $.

\end{document}